\documentclass[%
superscriptaddress,
amsmath,amssymb,
aps,
prl,
twocolumn,
]{revtex4-2}

\usepackage{graphicx}
\usepackage{dcolumn}
\usepackage{bm}
\usepackage{amssymb}
\usepackage{mathrsfs}
\usepackage{amsmath}
\usepackage{comment}
\usepackage{threeparttable}
\usepackage{lineno}
\usepackage{siunitx}
\usepackage{newtxtext}

\usepackage[driverfallback=dvipdfm]{hyperref}
\hypersetup{	colorlinks=true,
			linkcolor=blue,
			citecolor=blue,
			urlcolor=blue,
		}	

\begin{document}

\preprint{APS/123-QED}

\title{Search for the Pair Production of Dark Particles $X$ with $K_L^0 \to XX$, $X \to \gamma\gamma$}


\newcommand{\InstKorea}{\affiliation{Department of Physics, Korea University, Seoul 02841, Republic of Korea}}
\newcommand{\InstOsaka}{\affiliation{Department of Physics, Osaka University, Toyonaka, Osaka 560-0043, Japan}}
\newcommand{\InstChicago}{\affiliation{Enrico Fermi Institute, University of Chicago, Chicago, Illinois 60637, USA}}
\newcommand{\InstNTU}{\affiliation{Department of Physics, National Taiwan University, Taipei, Taiwan 10617, Republic of China}}
\newcommand{\InstSaga}{\affiliation{Department of Physics, Saga University, Saga 840-8502, Japan}}
\newcommand{\InstKyoto}{\affiliation{Department of Physics, Kyoto University, Kyoto 606-8502, Japan}}
\newcommand{\InstKEK}{\affiliation{Institute of Particle and Nuclear Studies, High Energy Accelerator Research Organization (KEK), Tsukuba, Ibaraki 305-0801, Japan}}
\newcommand{\InstYamagata}{\affiliation{Department of Physics, Yamagata University, Yamagata 990-8560, Japan}}
\newcommand{\InstJeonbuk}{\affiliation{Division of Science Education, Jeonbuk National University, Jeonju 54896, Republic of Korea}}
\newcommand{\InstJPARC}{\affiliation{J-PARC Center, Tokai, Ibaraki 319-1195, Japan}}
\newcommand{\InstNDA}{\affiliation{Department of Applied Physics, National Defense Academy, Kanagawa 239-8686, Japan}}

\InstChicago
\InstKorea
\InstOsaka
\InstNTU
\InstKEK
\InstKyoto
\InstJeonbuk
\InstJPARC
\InstNDA
\InstYamagata
\InstSaga
\author{C.~Lin}\InstChicago
\author{J.~K.~Ahn}\InstKorea
\author{J.~M.~Choi}\InstKorea
\author{M.~S.~Farrington}\InstChicago
\author{M.~Gonzalez}\InstOsaka
\author{N.~Grethen}\InstChicago
\author{Y.~B.~Hsiung}\InstNTU
\author{T.~Inagaki}\InstKEK
\author{I.~Kamiji}\InstKyoto
\author{E.~J.~Kim}\InstJeonbuk
\author{J.~L.~Kim}\InstJeonbuk
\author{H.~M.~Kim}\InstJeonbuk
\author{K.~Kawata}\InstOsaka
\author{A.~Kitagawa}\InstOsaka
\author{T.~K.~Komatsubara}\InstKEK\InstJPARC
\author{K.~Kotera}\InstOsaka
\author{S.~K.~Lee}\InstJeonbuk
\author{J.~W.~Lee}\InstKorea
\author{G.~Y.~Lim}\InstKEK\InstJPARC
\author{Y.~Luo}\InstChicago
\author{T.~Matsumura}\InstNDA
\author{K.~Nakagiri}\thanks{Present address: Department of Physics, University of Tokyo, Bunkyo, Tokyo 113-0033, Japan.}\InstKyoto
\author{H.~Nanjo}\InstOsaka
\author{T.~Nomura}\InstKEK\InstJPARC
\author{K.~Ono}\InstOsaka
\author{J.~C.~Redeker}\InstChicago
\author{T.~Sato}\thanks{Deceased.}\InstKEK
\author{V.~Sasse}\InstChicago
\author{T.~Shibata}\InstOsaka
\author{N.~Shimizu}\thanks{Present address: Department of Physics and The International Center for Hadron Astrophysics, Chiba University, Chiba 263-8522, Japan.}\InstOsaka
\author{T.~Shinkawa}\InstNDA
\author{S.~Shinohara}\thanks{Present address: KEK, Tsukuba, Ibaraki 305-0801, Japan.}\InstKyoto
\author{K.~Shiomi}\InstKEK\InstJPARC
\author{R.~Shiraishi}\InstOsaka
\author{S.~Suzuki}\InstSaga
\author{Y.~Tajima}\InstYamagata
\author{Y.-C.~Tung}\InstNTU
\author{Y.~W.~Wah}\InstChicago
\author{H.~Watanabe}\InstKEK\InstJPARC
\author{T.~Wu}\InstNTU
\author{T.~Yamanaka}\InstOsaka
\author{H.~Y.~Yoshida}\InstYamagata

\collaboration{KOTO Collaboration} \noaffiliation

\date{\today}
             
\begin{abstract}
We present the first search for the pair production of dark particles $X$ via $K_L^0\to XX$ with $X$ decaying into two photons using the data collected by the KOTO experiment. No signal was observed in the mass range of 40 -- 110~MeV/c$^2$ and 210 -- 240~MeV/c$^2$. This sets upper limits on the branching fractions as $\mathcal{B}(K_L^0 \to XX)$ $<$~(1--4)~$\times$~10$^{-7}$ and $\mathcal{B}(K_L^0 \to XX)$ $<$~(1--2)~$\times$ 10$^{-6}$ at the 90\% confidence level for the two mass regions, respectively.
\end{abstract}

\maketitle



Dark particle search is one of the major efforts in particle physics. The $s \to d$ quark transitions may result in more than one dark particle $X$ \cite{dark_pair_theory}. The signature of $K_L^0 \to XX$ with $X \to \gamma \gamma$ is unique because $K_L^0$ can directly couple to a pair of dark particles $X$, whereas $K^+$ requires an extra coupling to a Standard Model particle to conserve charge. Depending on the $X$ mass, the dark pair may appear in a $K_L^0$ decay but be kinematically forbidden in $K^+$ decays. The dark pair can be experimentally investigated if $X$ can promptly decay into two photons via a heavy quark loop.  To date, no experimental result has been reported on such decays.

The $K_L^0\to XX$ search was performed with the data collected by the J-PARC KOTO experiment \cite{KOTO_proposal, KOTO}. The 30-GeV protons hit a gold target, and secondary particles extracted at 16$^{\circ}$ from the proton beam line were collimated with a solid angle of 7.8~\SI{}{\micro sr} \cite{KL_beamline}. A 70-mm-thick lead block and a sweeping magnet were installed to eliminate photons and charged particles, respectively. The $K_L^0$ momentum distribution peaked at 1.4~GeV/c and was measured prior to the physics run by reconstructing the $K_L^0 \to \pi^0 \pi^+ \pi^-$ decay with the hodoscope system \cite{KL_momentum}. The $K_L^0$ flux at the entrance of the KOTO detector, 21~m downstream from the target, was 2~$\times$~10$^{-7}$ $K_L^0$ per proton on target. This was measured using the three $K_L^0$ decay channels: $K_L^0 \to 3 \pi^0$, $K_L^0 \to \pi^0 \pi^0$, and $K_L^0 \to \gamma \gamma$ \cite{KL_flux}.

\begin{figure*}
\includegraphics[width=0.99\textwidth]{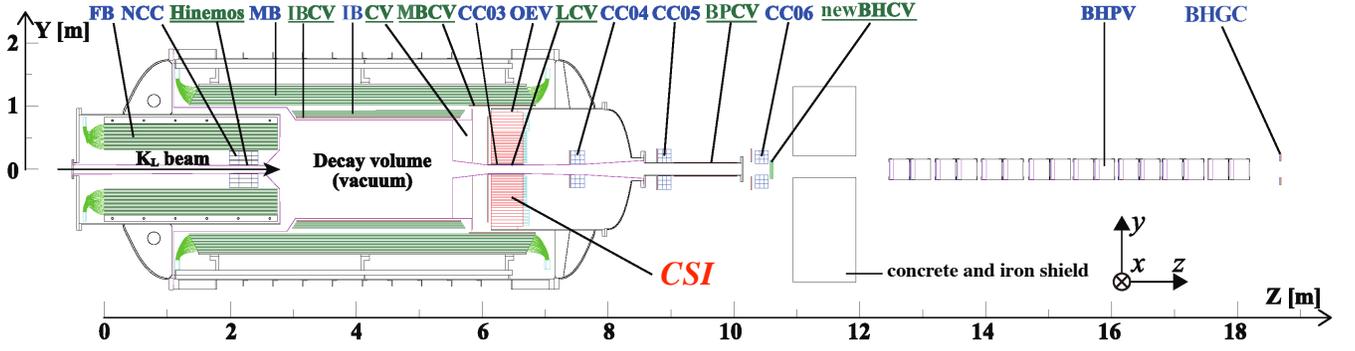}
\caption{\label{fig:koto_detector} Cross-sectional view of the KOTO detector. The names with (without) underline are dedicated to charged particle (photon) detection.}
\end{figure*}

Figure~\ref{fig:koto_detector} shows the schematic view of the KOTO detector. The $z$-axis lies along the beam center and points downstream. The energy and position of incident photons from $X$ decays were measured by a Cesium Iodide (CSI) calorimeter, which was a 1.9-m-diameter and 50-cm-long (27 $X_0$, where $X_0$ is the radiation length) cylinder with a 15~$\times$~15~cm$^2$ square beam hole at the center. Undoped Cesium Iodide crystals with a cross section of 2.5~$\times$~2.5~cm$^2$ (5.0~$\times$~5.0~cm$^2$) were stacked in the central (outer) region \cite{CSI}. The following detectors were used as veto counters to verify that there were no additional particles besides the four photons measured at the CSI. The lead-scintillator sandwich counters enclosing the decay volume (FB, MB, and IB \cite{IB}) and the photon veto counter at the outer edge of CSI (OEV) \cite{OEV} were used to detect extra photons. The collar counters (NCC and CC03-CC06) were made of undoped Cesium Iodide crystals and placed along the beam axis to detect escaping particles. The counters attached at the inner surface of IB and MB (IBCV and MBCV, respectively) were made of plastic scintillators to detect charged particles. The counter with two layers of 3-mm-thick plastic scintillators (CV) was used to detect charged particles hitting CSI \cite{CV}. The lead-aerogel sandwich counter (BHPV) \cite{BHPV} and the lead-acrylic sandwich counter (BHGC) were used to detect photons passing through the beam hole. Other detector components not mentioned above were not used in this analysis. The entire decay volume was kept at 10$^{-5}$~Pa to eliminate particle interactions with residual gas. Pulse shapes of the outputs from the detector were recorded with either 125 MHz or 500 MHz digitizers.
%

This measurement was based on the data collected in June 2018. The proton beam power was 51~kW, and during the one-month data taking, 1.1 $\times$ 10$^{19}$ protons hit the target. The $K_L^0\to XX$ data was collected using the following two-level trigger criteria: the first level trigger required that the energy sum in CSI was larger than 550 MeV without any coincident hit in NCC, MB, IB, CV, and CC03-CC06, and the second level trigger required four electromagnetic showers in CSI. The energy thresholds for veto were sufficiently higher than the thresholds used in offline analysis to avoid signal loss. Details of the data acquisition are available in Ref~\cite{DAQ}. 


Four momenta of $K_L^0$ and two $X$ particles were reconstructed as follows. First, crystals that had energy larger than 3 MeV within 71 mm of each other were grouped together to form a cluster. The cluster energy is defined as the sum of energy deposits in each cluster. The hit position (timing) was calculated by taking the average of crystal positions (timings) weighted by energy deposit in each crystal. Further information on the hit and energy resolution is available in Ref~\cite{CSI}. If $K_L^0$ traveled from the target to the center of energy (COE) in CSI and $X$ decayed into two photons promptly, the $K_L^0$ decay vertex ($Z_{vtx}$) was obtained by solving the following equation:
\begin{linenomath}\begin{align}
    M_{K_L^0}^2 = \sum_{i<j}^{4} 2 E_{\gamma_i} E_{\gamma_j} (1 - \cos{\theta_{\gamma_i \gamma_j}(Z_{vtx})}) ,
\end{align}\end{linenomath}
where $M_{K_L^0}$ is the nominal $K_L^0$ mass \cite{PDG2020}, $E$ is the photon energy, and $\theta_{\gamma_i \gamma_j}$ is the opening angle between $\gamma_i$ and $\gamma_j$ and a function of $Z_{vtx}$. The measured cluster energy is smaller than the incident photon energy due to the shower leakage, and the measured hit position is different from the point of incidence due to the finite size of the CSI crystals. By using the reconstructed incident photon angle, the energy and position of each photon were corrected. The $K_L^0$ vertex was then recalculated with the corrected photon energies and positions. The masses of the two $X$ particles were calculated for the three possible pairings and the one that had the reconstructed $X$ masses ($M_{X}$) closest to each other was selected. The difference between two reconstructed $M_X$ values ($\Delta M_X$) was required to be less than 10~MeV/c$^2$. The average of the two reconstructed $M_X$ values, defined as $\overline{M_X}$, is used to represent the $M_X$ of the event.


In order to ensure that the electromagnetic showers were fully contained in CSI, the hit position ($x$, $y$) of each photon was required to be inside the CSI fiducial region: $\min{(|x|,|y|)}>$~150~mm and ${\sqrt{x^2+y^2}}<$~850~mm. The distance between any two photon hits was required to be larger than 150~mm to ensure that the electromagnetic showers were isolated from each other. The timing difference between any two photon hits was required to be less than 3~ns to ensure that they were from the same $K_L^0$ decay. The $K_L^0$ energy was required to be larger than 650~MeV to eliminate loss due to the trigger requirements.

The signal acceptance and the background reduction were evaluated by Monte Carlo (MC) simulation using GEANT4 \cite{g4_1, g4_2, g4_3}. The simulated detector response was overlaid with the accidental hits induced by the beam, including extra particle hits from the beamline and pileup $K_L^0$ decays. The accidental hits were detected using the target monitor \cite{TMON} which issued triggers based on the secondary particles produced at the target reflecting the timing structure of the beam. 

One of the major background sources was the $K_L^0 \to 3 \pi^0$ decay with two missing photons. They could be missed in the veto counters due to their finite detection efficiency, or a photon hit could be hidden by another if they were too close to each other in the CSI such that the resulting electromagnetic showers were merged into one cluster (fusion cluster). A stringent threshold was applied to FB, NCC, MB, and IB in order to suppress $K_L^0 \to 3 \pi^0$ background by detecting extra photons. The size of a fusion cluster tends to be larger than the size of a single photon cluster. The cluster size was evaluated by the energy-weighted average of the distances between the crystals of a cluster and the hit position, and it was required to be less than 40~mm. The cluster shape was compared with a shape template of single photon hits, where the mean and the standard deviation of the energy deposit in crystals were provided for various incident photon angles and energies. The shape-$\chi^2$, which was the $\chi^2$ test calculated by comparing the shower shape to the template, was required to be less than 7 for all four clusters. A signal event would have all final-state particles hitting the CSI so the distance between the COE and the beam axis was required to be less than 50~mm. 

After applying all the selection criteria (cuts) described above, the majority of the remaining $K_L^0 \to 3\pi^0$ events had two fusion clusters from six final-state photons. The photon pair from each $\pi^0$ in the $K_L^0 \to 3 \pi^0$ decay are denoted by ($\gamma_1$, $\gamma_2$), ($\gamma_3$, $\gamma_4$), and ($\gamma_5$, $\gamma_6$). If $\gamma_1$ and $\gamma_3$ are merged together and $\gamma_2$ and $\gamma_4$ are merged together, the resulting two $M_X$ values would be $2 M_{\pi^0}$ and $M_{\pi^0}$, where $M_{\pi^0}$ denotes the nominal $\pi^0$ mass. This can be reduced by the $\Delta M_X$ cut. However, if $\gamma_2$ and $\gamma_3$ are merged together and $\gamma_4$ and $\gamma_5$ are merged together, the resulting $\Delta M_X$ may be small. The likelihood of an event induced by this mechanism is evaluated as follows. The six photon energies were explicitly solved by the following constraints:
\begin{linenomath}\begin{align}
 &2E_{\gamma_1}E_{\gamma_2}(1 - \cos{\theta_{\gamma_1 \gamma_2}}) = 2E_{\gamma_5}E_{\gamma_6}(1 - \cos{\theta_{\gamma_5 \gamma_6}}) , \label{eq:inv_mass} \\
    &2E_{\gamma_3}E_{\gamma_4}(1 - \cos{\theta_{\gamma_3 \gamma_4}}) = M_{\pi^0}^2 ,
\end{align}\end{linenomath}
where $\theta_{\gamma_i \gamma_j}$ is the opening angle between $\gamma_i$ and $\gamma_j$ calculated from the reconstructed $K_L^0$ vertex. Because Eq.~\ref{eq:inv_mass} should be $M_{\pi^0}$ for the $K_L^0 \to 3 \pi^0$ background with the double fusion, the double fusion difference $\Delta M^2$ ($\Delta M_{DF}^2$) defined below is expected to be small:  
\begin{linenomath}\begin{align}
    \Delta M_{DF}^2 = (M_{\gamma_1 \gamma_2} - M_{\pi^0})^2 + (M_{\gamma_5 \gamma_6} - M_{\pi^0} )^2 ,
\end{align}\end{linenomath}
where $M_{\gamma_1 \gamma_2}$ and $M_{\gamma_5 \gamma_6}$ are the reconstructed invariant-mass values calculated in the left-hand side and the right-hand side of Eq.~\ref{eq:inv_mass}, respectively. These calculations were performed for all the possible combinations of selecting two fusion clusters from the four clusters. The minimum of $\Delta M_{DF}^2$ among all the combinations was required to be larger than 1000 (MeV/c$^2$)$^2$; 96\% of the remaining $K_L^0 \to 3\pi^0$ background events were further removed.

\begin{figure}[h]
\includegraphics[width=0.48\textwidth]{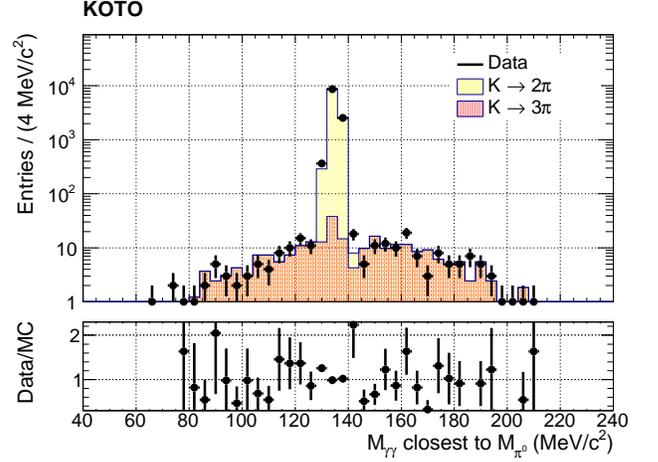}
\caption{\label{fig:norm} Distribution of the photon pair invariant mass that is closest to $M_{\pi^0}$ ($M_{\gamma\gamma}$ closest to $M_{\pi^0}$) after imposing all the cuts against the $K_L^0 \to 3 \pi^0$ background except for the $\Delta M_{DF}^2$ cut. The dots and the histograms indicate the data and the MC prediction, respectively.}
\end{figure}

The $K_L^0 \to \pi^0 \pi^0$ background is the special case of $M_X = M_{\pi^0}$, and therefore the $\Delta M_X$ was expected to be small. By requiring $\overline{M_X}$ to be outside of the $\pi^0$ mass window of 120~MeV/c$^2$--150~MeV/c$^2$, the reduction of 99.2\% was achieved. The remaining $K_L^0 \to \pi^0 \pi^0$ events had the wrong photon pairings because the correct photon pairings had a larger $\Delta M_X$. The correct photon pairings should have the reconstructed invariant mass close to $M_{\pi^0}$. Hence, the invariant masses of all the six possible photon pairings were calculated and the one that is closest to $M_{\pi^0}$ was used to suppress the $K_L^0 \to \pi^0 \pi^0$ background. As shown in Fig.~\ref{fig:norm}, the region of 120~MeV/c$^2$--150~MeV/c$^2$ is dominated by the $K_L^0 \to \pi^0 \pi^0$ decay and thus defined as a control region (CR). A signal was required to be outside of the CR.

The CR is used to normalize the MC to the data. After applying all the cuts except for the $\Delta M_{DF}^2$ cut, $N_{norm} =$ 11186 events were observed in data. 

\begin{figure}[b]
\includegraphics[width=0.48\textwidth]{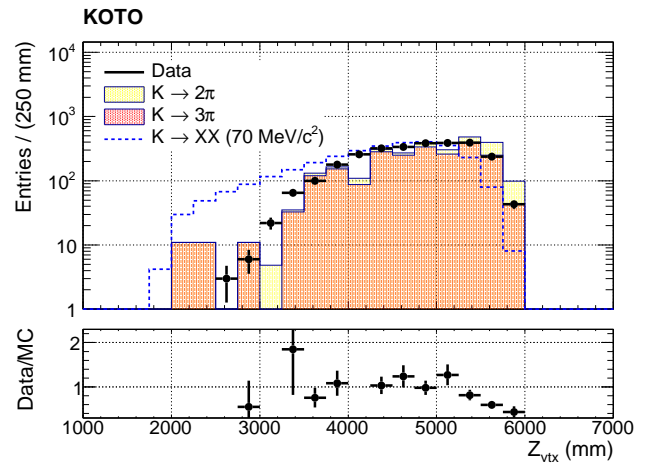}
\caption{\label{fig:klz_final} Distribution of $Z_{vtx}$ after imposing all the cuts except for the $Z_{vtx}$ cut. The dots and the histograms indicate the data and the MC prediction, respectively. The blue dashed histogram indicates the $K_L^0 \to XX$ distribution for $M_X$ $=$ 70 MeV/c$^2$ assuming the branching fraction of 5 $\times$ 10$^{-6}$. }
\end{figure}
\begin{figure}[b]
\includegraphics[width=0.48\textwidth]{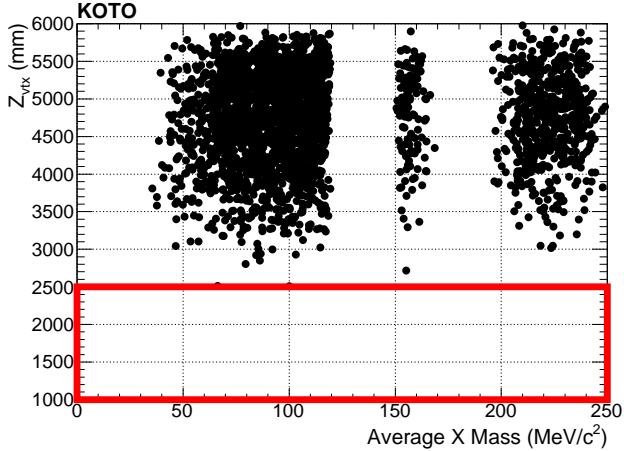}
\caption{\label{fig:xmass} Distribution of $Z_{vtx}$ versus average $M_{X}$ after imposing all the cuts except for the $Z_{vtx}$ cut. The thick red box represents the signal region. The dots indicate the data.}
\end{figure}

Figure~\ref{fig:klz_final} shows the $Z_{vtx}$ distribution after imposing all the cuts. The $K_L^0 \to 3 \pi^0$ background was suppressed in the upstream region. In order to obtain the most appropriate $Z_{vtx}$ cut, the $K_L^0 \to \pi^0 \pi^0$ decay was selected as signal and $K_L^0 \to 3 \pi^0$ decay was selected as background. The acceptance was defined as the number of events after imposing all the cuts except for the cuts against the $K_L^0 \to \pi^0 \pi^0$ background normalized to the number of $K_L^0$ mesons at the entrance of the KOTO detector in MC. The ratio of the $K_L^0 \to \pi^0 \pi^0$ acceptance ($A_{K_L^0 \to \pi^0 \pi^0}$) to the $K_L^0 \to 3 \pi^0$ acceptance ($A_{K_L^0 \to 3 \pi^0}$) was calculated for various $Z_{vtx}$ requirements. The $Z_{vtx}$ was required to be less than 2500 mm, which maximized the acceptance ratio. Figure~\ref{fig:xmass} shows the data distribution of $Z_{vtx}$ versus $\overline{M_{X}}$. The signal region was defined to be $Z_{vtx}$ $<$ 2500~mm and 0 MeV/c$^2$ $<$ $\overline{M_X}$ $<$ 250 MeV/c$^2$. The gaps ranging from 120 MeV/c$^2$ to 150 MeV/c$^2$ and from 160 MeV/c$^2$ to 190 MeV/c$^2$ were caused by the cut against the $K_L^0 \to \pi^0 \pi^0$ background and the $\Delta M_{DF}^2$ cut, respectively. After imposing all the cuts, no signal was observed. The number of $K_L^0 \to 3\pi^0$ and $K_L^0 \to \pi^0 \pi^0$ background events in the signal region was estimated to be (0.61 $\pm$ 0.61) and $<$~0.62 at the 90\% confidence level (CL), respectively. This was statistically consistent with the background prediction.


\begin{figure}[h]
\includegraphics[width=0.48\textwidth]{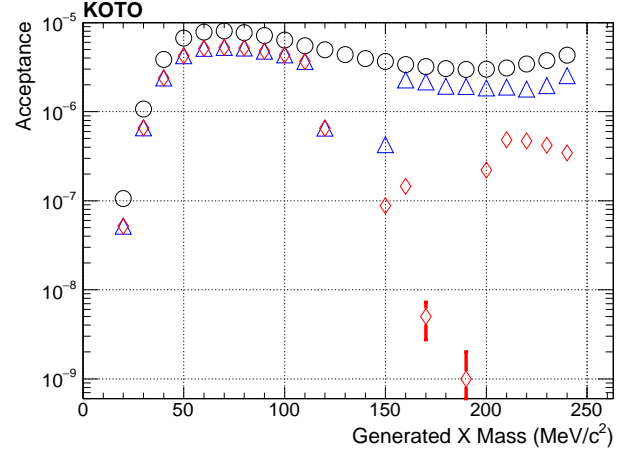}
\caption{\label{fig:KLXX_A} Signal acceptance versus $M_X$. The red diamond, blue triangle, black circular markers indicate the acceptance after imposing all the cuts, all but the $\Delta M_{DF}^2$ cut, and further excluding the cuts against the $K_L^0 \to \pi^0 \pi^0$ background, respectively.}
\end{figure}

The branching fraction was evaluated by the number of observed signal events multiplied with the single event sensitivity (SES), and the SES was calculated as
\begin{linenomath}\begin{align}
&SES = \frac{1}{A_{sig}} \times \nonumber \\
&\frac{A_{K_L^0 \to \pi^0 \pi^0} \mathcal{B}(K_L^0 \to \pi^0 \pi^0) + A_{K_L^0 \to 3 \pi^0} \mathcal{B}(K_L^0 \to 3 \pi^0) }{ N_{norm} }
\label{eq:ses},
\end{align}\end{linenomath}
where $A_{sig}$ is the acceptance evaluated by the $K_L^0 \to XX$ MC after applying all the cuts, and $\mathcal{B}(K_L^0 \to \pi^0 \pi^0)$ = 8.64 $\times$ 10$^{-4}$ and $\mathcal{B}(K_L^0 \to 3 \pi^0)$ = 19.52\% are the nominal branching fractions of $K_L^0 \to \pi^0 \pi^0$ and $K_L^0 \to 3 \pi^0$, respectively \cite{PDG2020}. The $K_L^0 \to XX$ was simulated for $M_X$ ranging from 10 MeV/c$^2$ to 240 MeV/c$^2$. The $\overline{M_X}$ of the $K_L^0 \to XX$ may differ from the generated $M_X$ due to the wrong pairing like in the case of $K_L^0 \to \pi^0 \pi^0$. The $\overline{M_X}$ cut of $\pm$ 10 MeV/c$^2$ of the $M_X$ to be examined was further required. Figure~\ref{fig:KLXX_A} shows the distribution of signal acceptance versus generated $M_X$. The mass region of 110 MeV/c$^2$--140 MeV/c$^2$ could not be examined due to the $K_L^0 \to \pi^0 \pi^0$ background. The mass region of 140 MeV/c$^2$--200 MeV/c$^2$ had a large signal loss introduced by the $\Delta M_{DF}^2$ cut. 

%
%

\begin{figure}[h]
\includegraphics[width=0.48\textwidth]{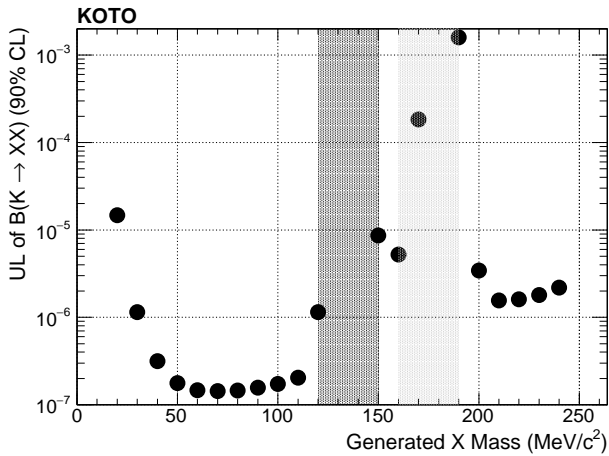}
\caption{\label{fig:KLXX_br} Upper limit (UL) of $\mathcal{B}(K_L^0 \to XX)$ for different $M_X$ at the 90\% CL. The dark (light) gray region indicates the exclusion by the cuts against the $K_L^0 \to \pi^0 \pi^0$ background (the $\Delta M_{DF}^2$ cut). }
\end{figure}

The systematic uncertainties of the SES are summarized in Table~\ref{tab:syst}. The uncertainties were estimated by the $K_L^0 \to \pi^0 \pi^0$ events in the $\overline{M_X}$ region of 125 MeV/c$^2$--145 MeV/c$^2$ after applying all the cuts except for the cuts against the $K_L^0 \to \pi^0 \pi^0$ background. The discrepancy between data and MC after imposing a cut was measured through the double ratio $r$:
\begin{linenomath}\begin{align}
 r = \frac{n_{\text{MC}}/\overline{n}_{\text{MC}}}{n_{\text{data}}/\overline{n}_{\text{data}}}    
\end{align}\end{linenomath}
where $n_{\text{MC (data)}}$ is the number of events after imposing all the cuts and $\overline{n}_{\text{MC (data)}}$ is the number of events after excluding one of the cuts. The deviation of $r$ from 1 is the uncertainty of a cut. The quadratic sums of those deviations of all the veto cuts, kinematic cuts, and shape-$\chi^2$ cut were quoted as the uncertainty of offline veto, kinematic selection, and shape-$\chi^2$, respectively. In particular, the $\Delta M_{DF}^2$ had the uncertainty of 0.4\% included in the kinematic selection uncertainty. The offline veto had the largest uncertainty. The signal loss caused by MB and IB in data largely differed from the MC prediction and the source remains unknown. The uncertainty from the MC statistics was calculated for different $M_X$ using Binomial statistics. The $M_X$ of 70 MeV/c$^2$ (170 MeV/c$^2$) had the smallest (largest) MC statistical uncertainty of 1.4\% (44.7\%). Because their signal acceptances differed by more than $\mathcal{O}(10^3)$, this resulted in a large variation in the MC statistical uncertainty. The evaluation of the online trigger uncertainty was based on the minimum-biased data, which had the trigger decision recorded but not applied. The loss after requiring the absence of online veto was quoted as an uncertainty. The uncertainty of the $\mathcal{B}(K_L^0 \to \pi^0 \pi^0)$ was obtained from the PDG \cite{PDG2020}. The uncertainties of other sources were smaller than those of the sources described above. In total, the statistical and systematic uncertainties were estimated to be 0.9\% and 14.3--46.9\%, respectively. An upper limit on branching fraction was set using Poisson statistics with the consideration of the systematic uncertainty fluctuation \cite{br_syst}. The upper limits on the branching fractions for different $M_X$ values are shown in Fig.~\ref{fig:KLXX_br}.

\begin{table}
\caption{\label{tab:syst}%
Summary of the systematic uncertainties in the SES.
}
\begin{ruledtabular}
\begin{tabular}{lc}
Source & Uncertainty [\%] \\
\hline
Offline veto & 13.0 \\
Kinematic selection & 4.7 \\
Shape-$\chi^2$ & 2.7 \\
MC statistics & 1.4 -- 44.7 \\
Online cluster-counting & 1.2 \\
Online veto & 1.0 \\
Geometrical & 0.1 \\
$\mathcal{B}(K_L^0 \to \pi^0 \pi^0)$ & 0.7 \\
\hline
\hline
Total & 14.3 -- 46.9 
\end{tabular}
\end{ruledtabular}
\end{table}

%
%


In conclusion, we searched for dark particle pairs produced in the $K_L^0$ decay by assuming that dark particles decayed into two photons promptly. Because no signal was observed, the branching fraction limits of $\mathcal{B}(K_L^0\to XX)<$~(1--4)~$\times$~10$^{-7}$ for 40 -- 110~MeV/c$^2$ and $\mathcal{B}(K_L^0 \to XX)<$~(1--2)~$\times$ 10$^{-6}$ for 210 -- 240~MeV/c$^2$ were set, respectively.


\begin{acknowledgments}	
	We would like to express our gratitude to all members of the J-PARC Accelerator and Hadron Experimental Facility groups for their support. 
	We also thank the KEK Computing Research Center for KEKCC and the National Institute of Information for SINET4. This material is based upon work supported by the Ministry of Education, Culture, Sports, Science, and Technology (MEXT) of Japan and the Japan Society for the Promotion of Science (JSPS) 
    under the MEXT KAKENHI Grant Number JP18071006 and the JSPS KAKENHI Grant Numbers JP23224007, JP16H06343, JP16H02184, JP17K05479, and JP21H04995, through the Japan-U.S. Cooperative Research Program in High Energy Physics; the U.S. Department of Energy, Office of Science, Office of High Energy Physics, 
	under Award Numbers DE-SC0009798; the Ministry of Education and the Ministry of Science and Technology in Taiwan under Grants No. 104-2112-M-002-021, 105-2112-M-002-013 and 106-2112-M-002-016; 
	and the National Research Foundation of Korea (2019R1A2C1084552 and 2022R1A5A1030700). Some of the authors were supported by Grants-in-Aid for JSPS Fellows.
\end{acknowledgments}	


\bibliography{myref}

\end{document}